Center for Economic Strategy and Competitiveness

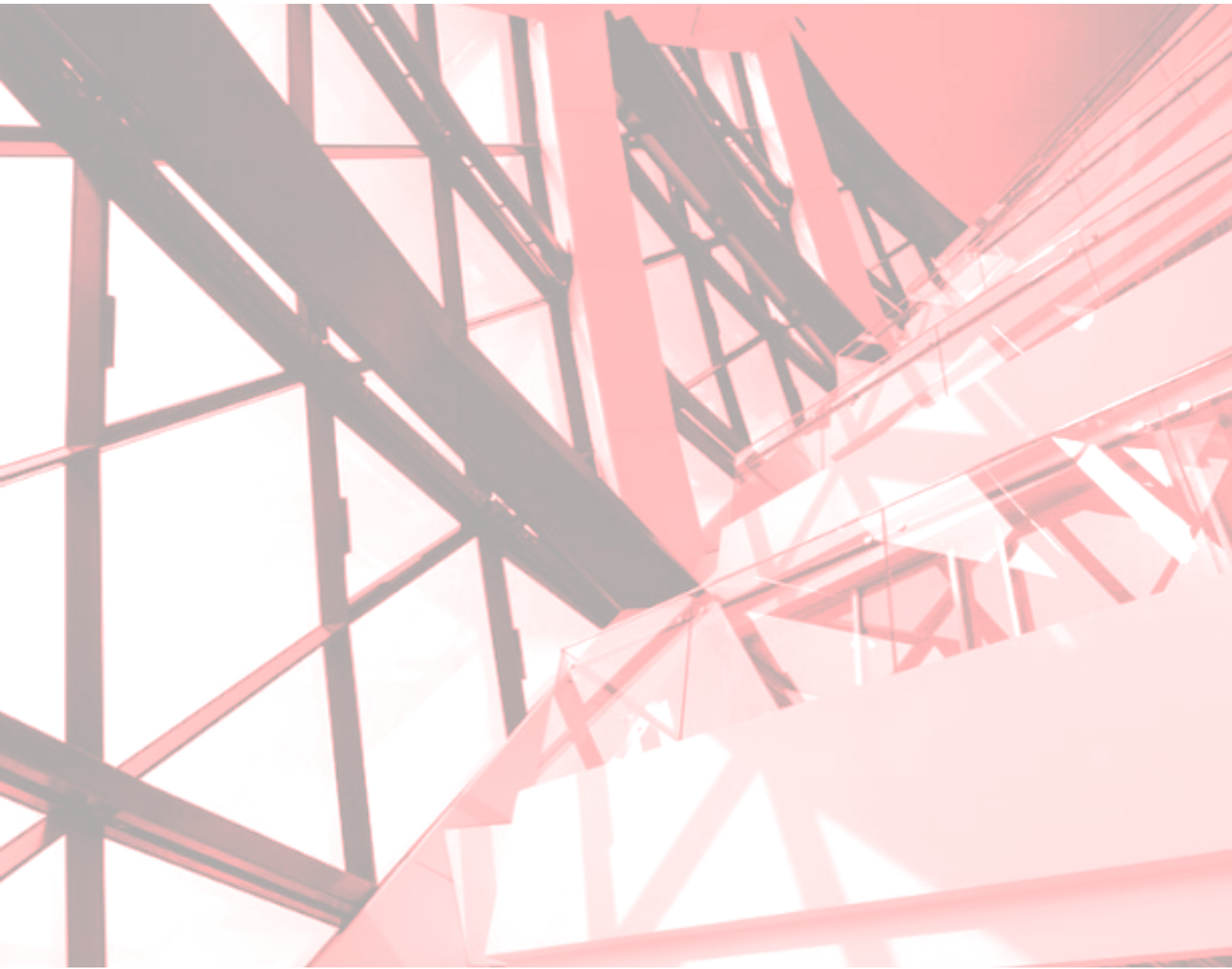

# Endogenous growth:
# Dynamic technology augmentation of Solow's model

**Murad Kasim**
mkasim@uni-sofia.bg
Sofia University "St. Kliment Ohridski", 2017

# Abstract

In this paper, I endeavour to construct a new model, by extending the classic exogenous economic growth model by including a measurement which tries to explain and quantify the size of technological innovation (*A*) endogenously.

I do not agree technology is a "constant" exogenous variable, because it is humans who create all technological innovations, and it depends on how much human and physical capital is allocated for its research. I inspect several possible approaches to do this, and then I test my model both against sample and real world evidence data. I call this method "dynamic" because it tries to model the details in resource allocations between research, labor and capital, by affecting each other interactively.

In the end, I point out which is the new residual and the parts of the economic growth model which can be further improved.

## Part I: Introduction

### 1. The problems of exogenous growth

The Solow-Swan model (1956)[1] is a simplified way to express economic growth:

$$Y = A \times L^{(1-\alpha)} \times K^{\alpha} \qquad (1)$$

In this model, the following variables are included:

*L* - **Labor**. Measured in counts of physical people which actively engage in work.

*A* - **Technology**. A variable explaining technology level, including innovation, ideas, research, know-how. Helps to create more output with the same *L* and *K*.

*K* - **Physical capital**. Measured in consumption goods.

*H* - **Human capital**. Used as a measure of the cumulative effect of education or training a person has received.

Physical capital (*K*) and labor (*L*) are considered as endogenous variables, contributing to output (*Y*) at a diminishing rate. Technology (*A*) is assumed as an endogenous constant. Later Mankiw, Romer and Weil[2] augmented the model by adding human capital (*H*) as a multiplier to labor (*L*):

$$Y = A \times HL^{(1-\alpha)} \times K^{\alpha} \qquad (2)$$

However, even the augmented model does not explain large variations in GDP per capita in different countries, based only on ***s*** (savings from capital which are reinvested), ***n*** (population growth rates) and *H* (quality of the human capital), and leaves this, again, to the so-called "residual", which is technology (*A*). This problem of Solow model's simplification becomes even more obvious as the share technological innovations continues to grow at present.

---

[1] Solow, R. (1956). *A Contribution to the Theory of Economic Growth*. The Quarterly Journal of Economics, Vol. 70, No. 1 (Feb., 1956), pp. 65-94. The MIT Press. Available at: http://www.jstor.org/stable/1884513. [Accessed 1 Aug. 2017]

[2] Mankiw, N. Romer, D. Weil, D. (1992). *A Contribution to the Empirics of Economic Growth*. The Quarterly Journal of Economics, Vol. 107, No. 2 (May, 1992), pp. 407-437. The MIT Press. Available at: http://www.jstor.org/stable/2118477. [Accessed 2 Aug 2017]

During the 1960s this imperfection has been pointed out by Kenneth Arrow (1962)[3], Hirofumi Uzawa (1965)[4] and Miguel Sidrauski (1967)[5]. Uzawa suggested a simpler model, in which evolution of A is determined by allocation of resources between a research sector and a final-goods sector.

Paul Romer (1986)[6], Robert Lucas (1988)[7], Sergio Rebelo (1991)[8] and Ortigueira & Santos (1997)[9] claimed growth is due to investment in human capital (*H*), which produced spillover effects.

---

[3] Arrow, K. (1962). *The Economic Implications of Learning by Doing*. The Review of Economic Studies, Vol. 29, No. 3 (Jun., 1962), pp. 155-173. The Review of Economic Studies Ltd. Available at: http://www.jstor.org/stable/2295952 [Accessed 2 Aug. 2017]

[4] Uzawa, H. (1965). *Optimum Technical Change in An Aggregative Model of Economic Growth*. International Economic Review, Vol. 6, No. 1. (Jan., 1965), pp. 18-31. International Economic Review. Available at: http://links.jstor.org/sici?sici=0020-6598%28196501%296%3A1%3C18%3AOTCIAA%3E2.0.CO%3B2-Y [Accessed 2 Aug. 2017]

[5] Sidrauski, M. (1967). *Rational Choice and Patterns of Growth in a Monetary Economy*. The American Economic Review, Vol. 57, No. 2, Papers and Proceedings of the Seventy-ninth Annual Meeting of the American Economic Association (May, 1967), pp. 534-544. Available at: http://www.jstor.org/stable/1821653 [Accessed 2 Aug. 2017]

[6] Romer, P. (1986). *Increasing Returns and Long-Run Growth*. Journal of Political Economy, Vol. 94, No. 5. (Oct., 1986), pp. 1002-1037. Journal of Political Economy. Available at: http://links.jstor.org/sici?sici=0022-3808%28198610%2994%3A5%3C1002%3AIRALG%3E2.0.CO%3B2-C [Accessed 2 Aug 2017]

[7] Lucas, R. (1988). On the mechanics of economic development. Journal of Monetary Economics Vol. 22 3-42. Elseiver Science Publishers B. V. (North-Holland). Available at: https://www.parisschoolofeconomics.eu/docs/darcillon-thibault/lucasmechanicseconomicgrowth.pdf [Accessed 2 Aug. 2017]

[8] Rebelo, S. (1991). Long-Run Policy Analysis and Long-Run Growth. Journal of Political Economy Vol. 99, No. 3 (Jun., 1991) pp. 500-521. The Journal of Political Economy. Available at: http://www.journals.uchicago.edu/doi/pdfplus/10.1086/261764 [Accessed 2 Aug. 2017]

[9] Ortigueira, S. and Santos, M. S. (1997). *On the Speed of Convergence in Endogenous Growth Models*. American Economic Review, 87(3), pp. 383-399. Available at: https://miami.pure.elsevier.com/en/publications/on-the-speed-of-convergence-in-endogenous-growth-models [Accessed 2 Aug. 2017]

## 2. Endogenous growth: Technology as an endogenously influenced function[10]

Therefore, I propose a new model, which treats technology as an endogenous variable. My approach is different by not only by considering human capital accumulation ($H$) but also physical capital's contribution ($K$), together with the introduction a depreciation model for human capital ($H$). I also include in my considerations the patent discovery rates, incentives provided for research and the size of active labor force or researchers available (which turns out to affect the model greatly).

I have structured my findings in three sections:

**Theoretical part:** I present and compare different scenarios for calculation of technology ($A$) by trying to make as less as possible assumptions. I extend the production function with size of labor and physical capital allocated to research. I model the spillover effects of human capital's accumulation ($H$) to technology ($A$). I also include depreciation of human capital ($H$) and a new link of physical capital ($K$) towards technology ($A$).

**Empirical part:** I test the scaling of my proposed models with sample and real world data sets. I try to determine which aspects of technology ($A$) affect output per capita ($Y/L$) most significantly. I also present the use of machine learning methods on how to distinguish anomalies in subsets of data.

**Closing thoughts on general growth models and their limitations:** I try to explain why growth models are imperfect by design and which are the areas left for further improvement.

In order to achieve all these difficult tasks, during our research we will also benefit from the service of our virtual assistant, Lee Chong, who has just returned to Singapore, after getting his Engineering degree in MIT. He is now employed at "ACME CORP - Singapore" and is asked which division he would like to join: Production department, where the engineers write code for future businesses, or the smaller R&D department, in which code is not being written, but existing algorithms are improved and new ones are created. Lee decides to go with the second, one which initially does not make his boss happy. Working in R&D means Lee will not be able to really "earn" money for the company, or at least not until he achieves his first patent. But the question is, will the company profit from hiring Lee?

---

[10] I acknowledge that I have drawn my inspiration from Romer's paper on the topic (1990), however my approach on interpreting and modeling the problem are very different. I try to quantify some of the simplifications Romer has made. The main difference is I try to not focus so much on the division of human capital ($H$) but its combination with physical capital in order to allocate their product to research or production. Also, I use the terms "design", "patent" and "idea" interchangeably, because to my knowledge ideas cannot be measured in any other reliable way.

## Part II: Theoretical considerations

### 1. Overview of existing theoretical definitions on technology (A)

During the years, there have been different definitions of technological advancement (*A*). Solow (1956) treats it as global exogenous good, available to all countries. Shell (1967)[11] defines it as a public input, provided by the government. However, I do not consider these realistic enough, as even if technology was provided by an external party "for free", it would still cost time for labor force to learn it, and then to re-adapt the existing processes to it.

Arrow (1962) assumes that an increase in capital (*K*) leads to increase in knowledge as well, due to "learning by doing", however he still treats knowledge and technology (*A*) as public good.

Romer (1987, 1990) presents two models, which includes private, maximizing behavior's role in generating technological change, by introducing patent prices, monopoly options and human capital division, by referencing to the consumption goods theory of Dixit and Stiglitz (1977)[12] and Judd (1985)[13]. He models technology (*A*) as different "recipes" with which consumption goods (*K*) can be combined with labor (*L*), to produce output (*Y*). He also acknowledges that research has a cost: "Once the cost of creating a new set of instructions has been incurred, the instructions can be used over and over again at no additional cost. Developing new and better instructions is equivalent to incurring a fixed cost", however, he does not include this in his calculation for simplicity purposes (this is one of the points which I address in my model).

### 2. K and L: Physical capital and Labor force

Romer suggests that any employee (*L*) can choose to participate by either investing his human capital *(H)* to work in production of final output *(Y)* (which generates output for the current period *t*) or in human capital accumulation for technology *(A)*[14] (which generates output in future period *t + x*, where *x* is the amount of periods needed to reach the break-even point from the investment in research):

$$H = H_Y + H_A \qquad (3)$$

---

[11] Shell, K. (1967). *A Model of Inventive Activity and Capital Accumulation*. Essays on the Theory of Optimal Economic Growth. Chapter IV, pp. 67-85. Cambridge, Massachusetts: MIT Press.
[12] Stiglitz, J. and Dixit, A. (1977). Monopolistic Competition and Optimum Product Diversity. American Economic Review. Vol. 67. Available at: https://doi.org/10.7916/D8S75S91 [Accessed 2 Aug. 2017]
[13] Judd, K. (1985). Redistributive taxation in a simple perfect foresight model. Journal of Public Economics. Vol. 28 Issue 1 Oct. 1985, pp. 59-83. Available at: https://www.sciencedirect.com/science/article/pii/0047272785900209 [Accessed 2 Aug. 2017]
[14] Romer, P. (1990). *Endogenous Technological Change*. Journal of Political Economy. 1990, vol. 98, no. 5, pt. 2. The University of Chicago. p. 85. Available at: http://pages.stern.nyu.edu/~promer/Endogenous.pdf [Accessed 2 Aug. 2017]

However, he does not acknowledge the role of physical capital consumption, by accepting the following simplification of nonrivalry:

"For simplicity, the arguments here will treat designs as idealized goods that are not tied to any physical good and can be costlessly replicated, but nothing hinges on whether this is literally true or merely close to being true."

While I do agree with a simplification, which would ignore the cost of the paper, on which the design is printed or multiplied, the capital which I would like to focus on is the one "invisibly consumed" until the paper with the design is created.

But let's go back to our assistant Lee. His company "ACME CORP - Singapore" is famous around the world, thanks to it's unique human capital of intelligent and young engineers. However, would those engineers even consider to do research for the company, if it wasn't for its large and quiet testing labs, free hardware and infinitely "free" testing samples? While Lee has been investing mostly his human capital ($H$) only in the research, he is happily unaware of the research equipment costs, or even the cost of him not producing direct output for the present. Therefore, it costs capital ($K$) to invest in the creation of a new design ($A$), since it implies an opportunity cost. And then again it costs time and resources for the firm to apply the new design in their process. By conducting research in order to improve technology ($A$), certain portion of the capital ($K$) and labor ($L$) need to be allocated to research and development. Thus, investing in research to improve technology ($A$), actually decreases the capital ($K$) and labor ($L$) for immediate output. And this is why Lee's boss is unhappy he chose to work in the R&D department.

While Romer includes interest rate in his model, as a way of measuring how 'costly' it would be to invest in research at present, in order to benefit from it in future, I have chosen to follow a different approach, by dividing physical capital ($K$) and labor ($L$) to production and research parts:

$$K = K_{production} + K_{research} \quad (4)$$
$$L = L_{production} + L_{research} \quad (5)$$

**3. H: Human capital:**

Romer suggests human capital ($H$) is to be divided to two parts: research and production. In my model, I cannot consider how "one can be trained or study for innovation or production", so I express human capital ($H$) in its classic form. One might ask why human capital ($H$) cannot be infinitely effective, just like technology ($A$) in cutting-edge growth. I observe that as people become older, at first their human capital ($H$) increases steeply, then it stagnates and starts to depreciate slowly with age. No worker ($L$) is able to preserve his human capital (*H*) beyond his mortal life. Even worse, people start to underperform as they get older and after the person dies, the human capital (*H*) is lost (with some part of it, hopefully, transferred to the permanent technology (*A*) during his lifetime - this link is modeled with formulas (10) and (11). Therefore, I think it is more realistic to model human capital by adding a slight depreciation factor to it. I introduce a log-normal distribution function in order to model more realistically human capital's (*H*) following attributes: until a certain critical inflex point (which can be a certain age or IQ), workers benefit exponentially from increase in their human capital (*H*). Beyond this inflex point, they start to experience diminishing returns, and even slow decrease in their current productivity.

I provide two examples here to make my point easier to understand:

An university student, who excessively studies (not to be confused with doing research!) until his early thirties without actually putting to practice his learned skills at active labor, thus producing no output. He is well beyond his human capital's (*H*) inflex point and will need to recoup his extra time put in study (while hoping for a higher salary), unlike his peers, who engaged in the labor market earlier.

The second example is with one of the professors of Lee in MIT, named Mr. Stein, who is 79 years old. Lee remembers that while he enjoyed his professor, he often noted that he tends to forget some facts due to his age, together with some of his knowledge already being outdated.

Thus, we need to add to human capital (*H*) a depreciation rate $d$:

$$H^d \quad (6)$$

With the depreciation rate $d$ being defined recursively as:

$$d = \sqrt{\frac{lnH}{e^{lnH}}} \quad (7)$$

$d$ applies is log-normal distribution function which models the aforementioned anomalies with human capital (*H*) as it grows:

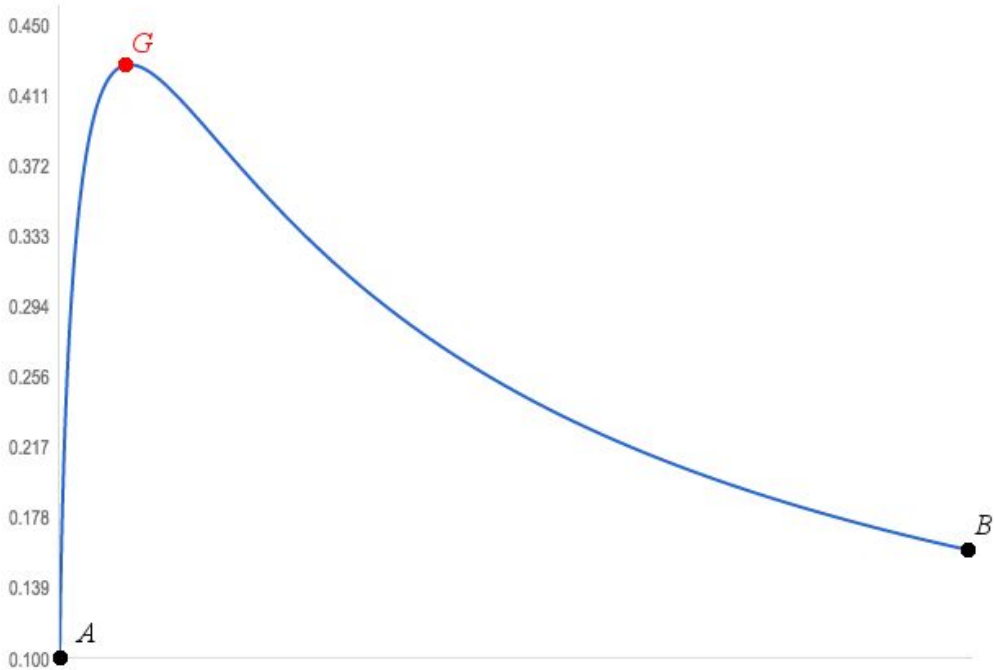

**Graph 1:** *d:* Human capital depreciation rate. Investing in human capital is initially exponential, then becomes subject to diminishing returns, then is slowly counterproductive after reaching an inflex point *G*. $\widehat{AG}$ is the optimal human capital accumulation (with subject to diminishing returns). $\widehat{GB}$ is the ineffective human capital accumulation.

With depreciation rate (*d*), human capital (*H*) has a critical inflex point, beyond which the effectiveness of investing in human capital depreciates. Below is the application of the function to sample data:

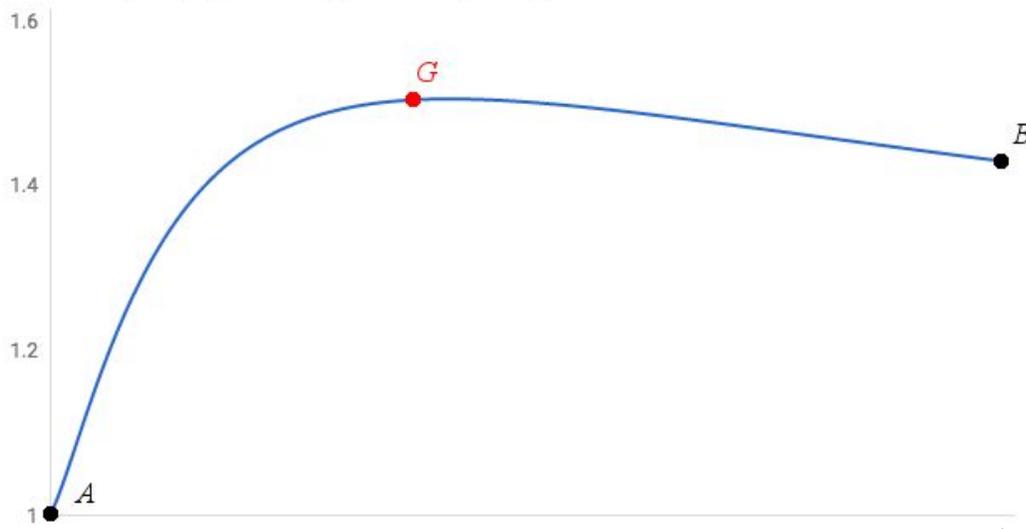

**Graph 2:** Human capital depreciation applied to data: The function models the diminishing ($\widehat{AG}$) and eventually, negative returns ($\widehat{GB}$) of intensive spending on human capital accumulation.

## 4. A: Technology, ideas, know-how

Another important effect is the cost of introducing or learning a new technology. It takes time and resources to produce each new innovation or patent. If Lee needs to learn a new library, in order to work with the new software, he is going to lose some days to learn it. However, it must be noted, that thanks to his higher human capital ($H$), he is able to learn it quicker (since he has more experience). Thus, we can point that there should be a relation between human capital ($H$) and costs of introducing a new technology ($A$). In order to model technology ($A$) reliably, I have tested few alternatives. The first one is suggested by Tabarrok and Cowen (2013)[15]:

$$A_1 = Population \times Incentives \times Ideas\ per\ hour \qquad (8)$$

The problem with this version is that it makes no sense to represent ideas per hour, as it does not change the model much. Another problem is that it takes as a measure the total population count, while Romer (1990) points out his error in his previous paper (1987) that it is more accurate to measure the human capital devoted to research instead. But perhaps the worst part is, if you try this formula with real data, it tends to scale too much, primarily due to Population. If Population is replaced with percent of population engaged in research (refer to **Appendix 2**), then it has better results, but it still overscales slightly, while introducing the problem of not taking the total population in account (which again leads to some distortions in very big or small countries). It also lacks human capital ($H$), since a more educated labor force ($L$) would be able to grasp technology ($A$) faster. Below, trying to avoid the aforementioned problems, I offer a better way of quantifying technology ($A$):

$$A_2 = 1 + (Researchers \times Incentives \times Ideas\ per\ capita) \qquad (9)$$

However, this model has a fundamental weakness: It does not scale with human capital ($H$).
How can a country with a low human capital ($H$) benefit from latest technological innovations ($A$)? Who is going to harness this knowledge, without education? The better trained the labor force ($L$) is, the faster they can develop new or embrace existing technologies ($A$). This is why, we are also adding a relation of $H$ to $A$.

---

[15] Tabbarok, A. and Cowen, T. (2013). Growth, Capital Accumulation, and the Economics of Ideas: Catching Up vs. Cutting Edge. *Modern Principles of Economics*. 2nd Edition. Worth Publishers. United States of America. p. 529.

So technology accumulation, augmented with the level of human capital becomes:

$$A_3 = 1 + \frac{(Researchers \times Incentives \times Ideas) \times Human\ capital\ level}{Labor\ force\ size} \quad (10)$$

The last model includes the full benefit from incentives and government spending on research. It scales well when simulated with sample data (refer to **Experiment 1** and **2**). My simulations with sample and real world data show the third model ($A_3$) to be the most stable, both with rich and poor countries, without any significant distortions (at least according to findings so far). Expressed in a stricter mathematical notion:

$$A = 1 + \frac{(P \times L_{research} \times K_{research}) \times H}{L} \quad (11)$$

Where $P$ is patent applications (ideas)[16], $L$ is size of labor force available, $L_{research}$ is labor force engaged in research and $K_{research}$ is capital size spent on research and $H$ is human capital level.

After comparing models $A_2$ and $A_3$ with sample data, the results is they differ only at their speeds of scaling (influenced via human capital ($H$)). Both of them display the same pattern:

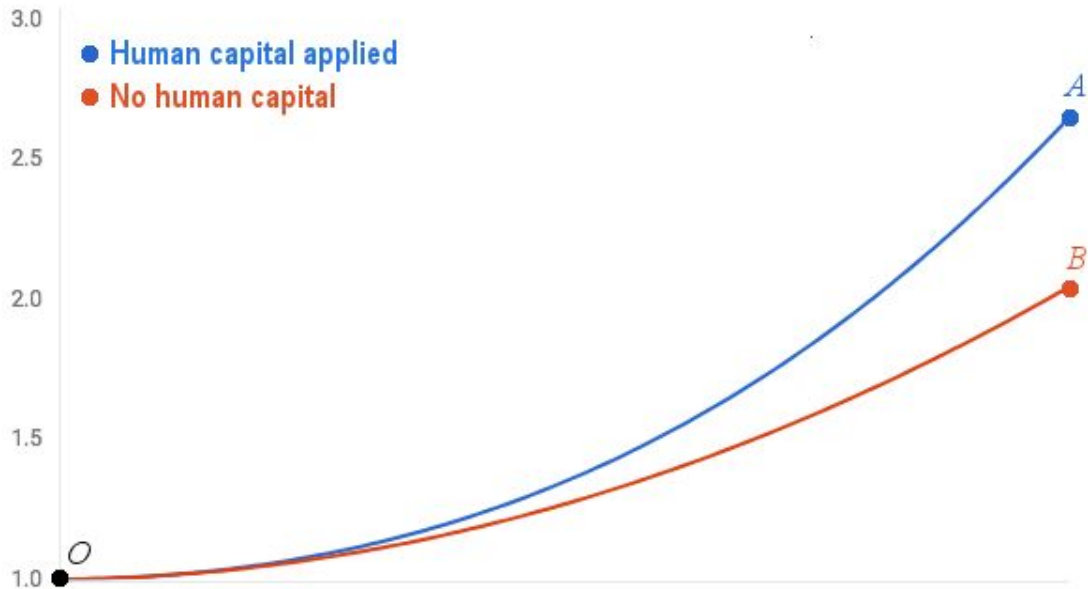

**Graph 3:** Comparison of technology functions growths.

---

[16] I choose to use patent applications, instead of approved patents, since an idea can be rejected by the patent office. The actual number of ideas should be even larger, since not all researchers and inventors patent their ideas.

Both models $A_2(\widehat{OB})$ and $A_3(\widehat{OA})$ are scaling decently, with $A_2$ scaling at a slower rate than model $A_3$, since it scales also with human capital (*H*). The latter suggests with higher levels of human capital (*H*) better growth of technology (*A*) can be achieved. But does a critical level of innovation and research exist? In the empirical section, my results suggest that underdeveloped countries cannot start to benefit from technology (*A*) until a critical mass of human capital (*H*) is accumulated (see **Experiment 6**), together with reaching the breakeven point period, after starting to invest in research (see **Appendix 7**).

### 5. Endogenous growth: Formulating the complete model

So, let's inspect again the classical production function again:

$$Y = A \times (H \times L)^{(1-\alpha)} \times K^{\alpha} \qquad (12)$$

After applying the above models of each variable, the dynamic endogenous growth function becomes:

$$Y = (1 + \frac{(P \times L_{research} \times K_{research}) \times H}{L}) \times H^d \times (L - L_{research})^{\alpha} \times (K - K_{research})^{\alpha}$$

$$(13)$$

## Part III: Description of methods, data and tools used

In this section I describe which methods and tools I use in testing the model against data evidence in the empirical section.

**1. Model assessment methods:**

The following variables and methods will be used to determine the fit of collected data and if the compared models are better or worse than their alternatives:

1. *t*-**statistic**: Departure ratio of a parameter, which is estimated from its notion value and its standard error. We will assume a critical value of 1.97[17], meaning that it has 95% chance of being statistically significant.

2. **p-value:** Identifies evidence for or against the null hypothesis. A small value (typically p≤ **0.0001**) indicates strong evidence against the null hypothesis and a large value ( p> **0.05**) indicates weak evidence against the null hypothesis (so the null hypothesis cannot be rejected). For ease of reading, when the value was <=0.0001, the exponential form of the actual number was hidden. Our selected critical value for lack of support on the null hypothesis will be p<0.05

3. **Schwarz criterion:** Also known as **Bayesian information criterion**, **the Schwarz criterion** is a used for model comparison among different models. The model with the lower value is identified as better.

4. **Akaike criterion:** Another criterion for model comparison (which is an alternative of Hannan-Quinn) which provides measure on the information lost about the constructed model, in order to represent the data. The model with lower value is preferred.

5. **Adjusted $R^2$** - The explanatory power of the regression (explained variation to total variation ratio). Interpretation[18]: worst: 0, best: 1. It is to be noted, that since in the current paper we will not filter our country data by groups (Oil countries, developing countries and etc.) the data set will contain more "contrasting" data, which will result in imperfect fits, leading to overall lower level of R-squared, but with the tradeoff of testing more universally applicable theories.

6. **Log-likelihood**: Cannot be used standalone as measure of good fit, but as subjective measure, the higher value, the better the model.

7. **Sum of squared residuals**: A measure of deviations between predicted and actual values. A smaller value, indicates a better fit, thus better model.

---

[17] http://itl.nist.gov/div898/handbook/eda/section3/eda3672.htm
[18] Windmeijer, F and Cameron, C. *An R-squared measure of goodness of fit for some common nonlinear regression models*. Journal of Econometrics. Volume 77, Issue 2, April 1997, Pages 329-342. Available at: https://www.sciencedirect.com/science/article/pii/S0304407696018180 [accessed May 5, 2017]

8. **Standard error of regression**: Smaller the values indicate that all data points are closer to the trendline (therefore, with less prediction errors). <u>Smaller values</u> provide evidence for a better model.

**2. Software tools used:**
- Data is stored in tables via Google Sheets[19].
- Regression analysis is done via Gretl[20].
- 2D graphs are visualized via Google Charts[21].
- The online service "Blockspring[22]" has been used for the machine learning algorithms calculation (K-Means clustering[23]).
- 3D data plots are visualized via Graph3D JavaScript library.[24]

**3. Variable building:**
Error correction methods have been applied for missing values of GDP spending on research. Derivative variables were built only when all data was present with no exceptions, with the entire country being excluded if at least one of the needed variables was missing, resulting in a clear set of 60 countries (see **Appendix 4**). Below are all the variables which were gathered or derived from the original variables:

**Variables descriptions & transformations:**
**P** - Total population of country
**L** - Active labor force of country (percentage of population aged between 15 and 64 years)
**A** - Level of dynamic technology. Calculated by the model of formula (11):
**Y** - Output (A production function of physical and human capital, labor and level of technology)
**Y/L** - Output per capita. Output (**Y**) divided by labor (**P**)
**h_1** - Human capital with average years of **secondary** education.
**n** - Labor force growth rate
$s_k$ - Capital savings rate. Derived from five year averages of **I/Y***, starting with 1965-1970 and finishing with 2005-2009 (with year 2010 not being included) and converted to percentages.
**δ** - Capital depreciation rate (assumed as constant: 0.02)

---

[19] Available at: https://www.google.com/sheets/about [Accessed 2 Aug. 2017]
[20] Available at: http://gretl.sourceforge.net/ [Accessed 2 Aug. 2017]
[21] Available at: https://developers.google.com/chart/ [Accessed 2 Aug. 2017]
[22] Available at: https://www.blockspring.com/ [Accessed 2 Aug. 2017]
[23] A detailed description of how clusterization and K-means clustering works is available here: http://www.mit.edu/~9.54/fall14/slides/Class13.pdf
[24] Available at: https://github.com/almende/chap-links-library/tree/master/js/src/graph3d [Accesed 2 Aug. 2017]

**α** - Capital's share income (assumed as 0.3333, based on recommendation of the RMW paper)

**Pa** - Patent applications (not to be confused with approved patents).

**Pc** - Ideas per capita. Calculated by dividing patent applications (**Pa**) to **L** (Labor).

**Ph** - Ideas per hour. Calculated by dividing patent applications (**Pa**) to 8760 (total hours in one year).

**Hd** - Human capital, with its depreciation rate applied. Calculated by the formulas (6) and (7).

**GDP R&D** - Percentage of country GDP allocated to research. Raw values were provided in float (Example: 2.23%) and were converted to actual percents (0.023).

**ln** - All variables have been transformed using the natural logarithm (based on Euler's number

*e* = **2.718281828459**) in order to achieve an abstraction from their absolute values and focus on their

**4. Data sets:**

**4.1. Real world data** (refer to **Appendix 2**):

    **1. Penn World Table**, which contains data for GDP and population related statistics.
The data set contains 190 countries, of which China 2 and Zaire has been excluded since CIV is the data for China and Zaire has existed only during the 70s of the 20th century and instead the full data for China and Congo has been considered, leaving the set at 188 countries. Country codes from ISO 3166-1 alpha-3 format have been converted to full names.

    **2. Barro-Lee Data on Education,** which contains country statistics on education attainment.
Data was present for the following countries, but no data was available in the other databases, so they were not added: Myanmar, Democratic Republic of the Congo, Gambia. Data for "Reunion" has been excluded.

    **3. Worldbank: World Development Indicators**. Data for patent applications by country's residents per year has been used from "Patent applications, residents". Country's spending on research has been fetched from "Research and development expenditure (% of GDP)". Due to many missing entries of data the following indicators could not be fetched: Researchers in R&D (per million people, Technicians in R&D (per million people).

    The final data set, after merge and sanitisation, has resulted in complete set of 60 countries:
Algeria, Argentina, Australia, Austria, Belgium, Bolivia, Brazil, Canada, China, Colombia, Costa Rica, Cyprus, Denmark, Ecuador, Egypt, El Salvador, Finland, France, Greece, Guatemala, Honduras, Hong Kong, Iceland, India, Indonesia, Iran, Ireland, Israel, Italy, Japan, Luxembourg, Malaysia, Mauritius, Mexico, Morocco, Netherlands, New Zealand, Nicaragua, Norway, Pakistan, Panama, Paraguay, Peru, Philippines, Portugal, Romania, Singapore, South Africa, South Korea, Spain, Sri Lanka, Sweden,

Switzerland, Thailand, Turkey, Uganda, United Kingdom, United States of America, Uruguay and Zambia.

Data has been filtered at 5 year intervals (1965, 1970, 1975, 1980, 1985, 1990, 1995, 2000 and 2005) in order to be able to run regressions involving capital savings rate and human capital (education attainment) The cleaned up files and original sources have been attached in the **Appendix 4**, with data collected and summarized for the period between 1960-2009.

**4.2. Sample data set:**

I have also generated a data set with sample, linearly increasing data. The purpose is to simulate a complete data set, without missing variables, in order to test the behavior of my models for which real life data was not available at present, due to incomplete reporting[25]. I hope in future those models can be ran with real life data as well. All variables in this data set are scaling linearly, with constant growth rates.

---

[25] The particular variables are amount of researchers per country and GDP percentage spending on research (prior year 1995).

**Part IV: Empirical tests and results**

In my tests, I start with the simplest model, and then I add regressors one by one, in order to confirm if the model becomes more accurate. In the end, I test with all variables of the model included.

*Experiment 1:* **Comparison of growth with constant and dynamic technology augmentation**

In this experiment, we will compare Solow's classic growth ($\widehat{OB}$) function against our dynamically augmented model ($\widehat{OA}$). Both models are using same linearly increasing values for human capital (H), but the dynamic model is allocating a percentage of physical capital (4%) and labor (0.025%) to research (refer to **Appendix 3**).

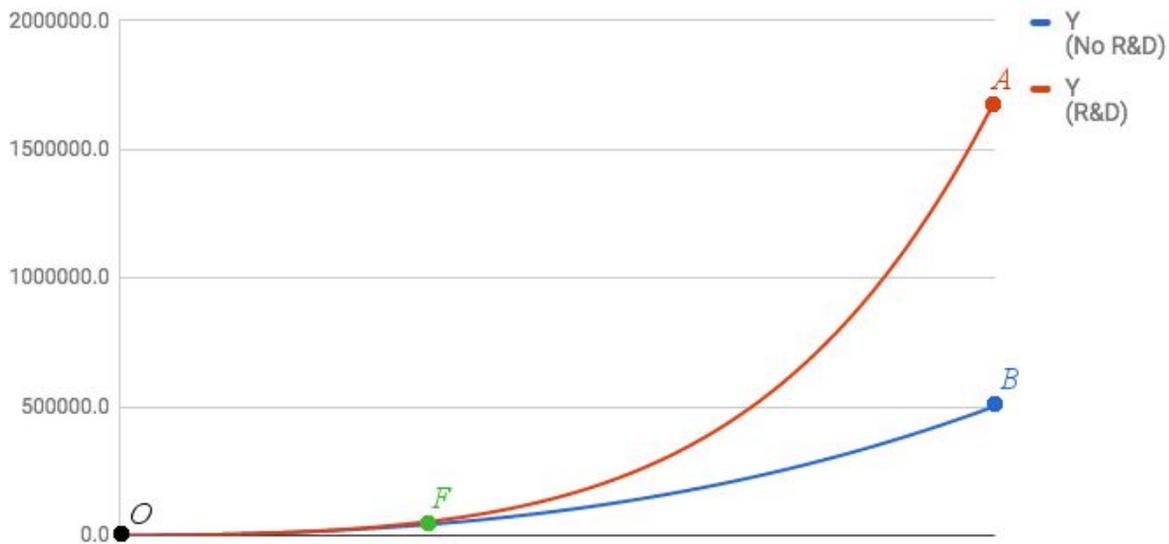

**Graph 4:** Growth function (*Y*): With and without dynamic technology augmentation (*A*)
**Data source:** Appendix 3

The augmented output ($\widehat{OA}$) initially scales slowly, due to increased costs on research. As seen on the zoomed **Graph** 5, between points *O* and *F*, output ($\widehat{OA}$) with research is actually less than output without research ($\widehat{OB}$) (since the investments in research do not produce output immediately). However after point *F* (which is the break-even point of returning all investments of research), the output function with research ($\widehat{OA}$) grows faster, thanks to the increased efficiency of applying the new technologies.

If we dive in in the data (**Appendix 3**), the model suggests that richer countries benefit more from research (exponentially in the long run). Also the diminishing returns of capital (*K*) is being reduced, confirming the original Solow model's assumption that technology (*A*) raises steady state directly.

However, all of this comes at cost. As seen later in **Experiment 5**, investing in technology (*A*) function requires a "critical mass" to be reached before starting to be beneficial.

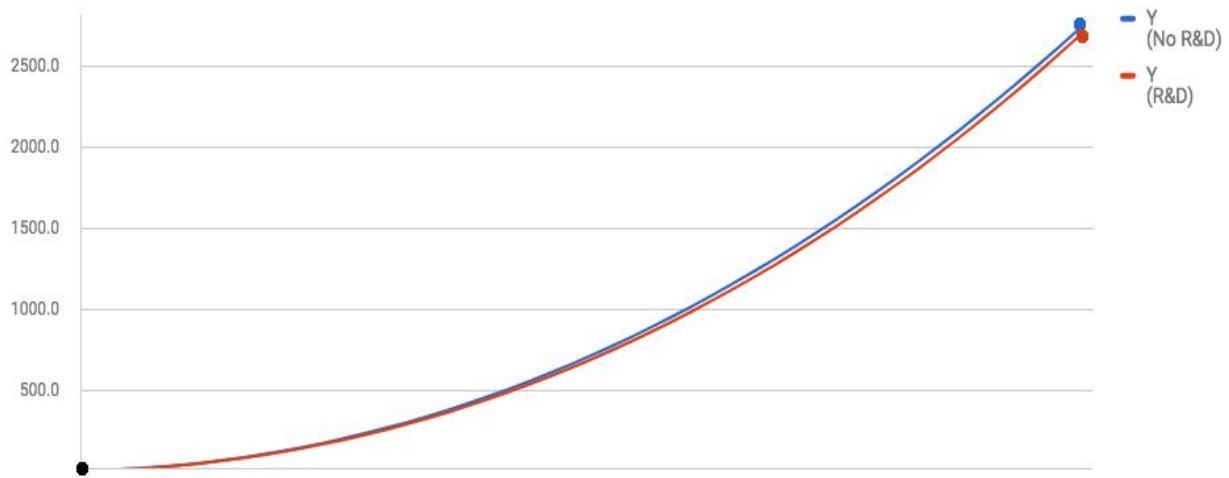

**Graph 5:** Growth function (*Y*): With and without application technology augmentation (*A*): Zoomed
**Data source:** Appendix 3

The model also implies (**Appendix 3**), the lower the technology advancement (*A*) and human capital (*H*) is, the more periods of ***t*** are required until output (*Y*) with technology augmentation $\widehat{OA}$ reaches its break-even point ***F*** and becomes profitable. So, while investment in immediate output production provides immediate result, for greater long term growth, it is better to invest in human capital (*H*) accumulation and research (*A*).

*Experiment 2:* **Test of dynamic technology model (*A*) with real world data**

In this experiment, I test the stability of my dynamic model for technology (*A*) (formula (11)) against real world data by using first classical and then human capital augmented Solow model.

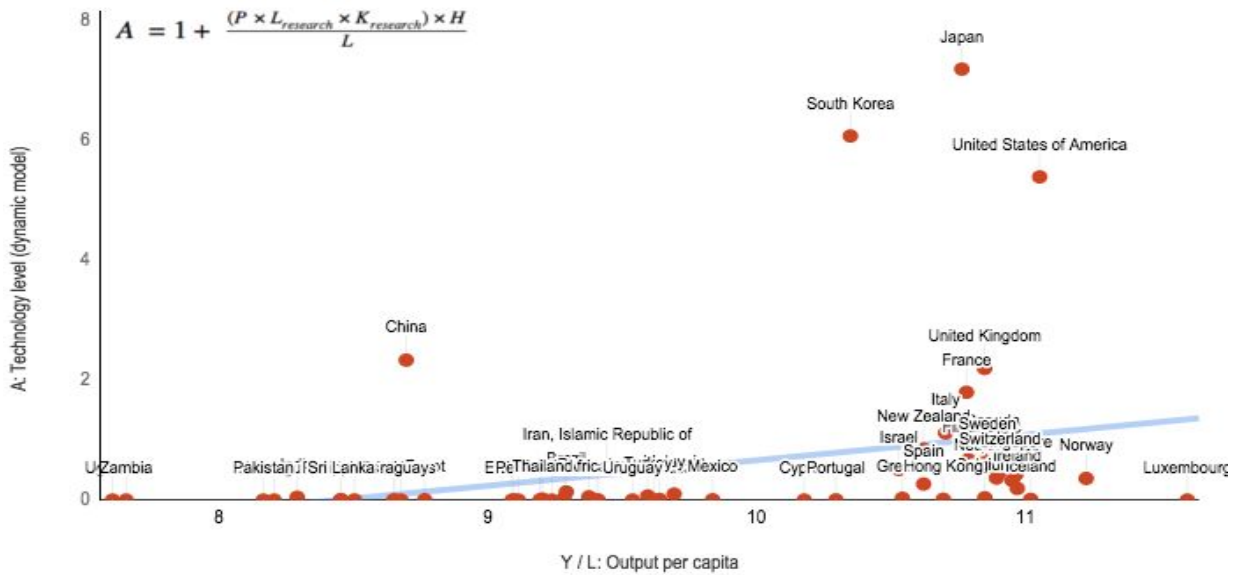

**Graph 6:** Output per capita (*Y*) vs dynamically augmented technology (*A*) levels. Logs values used.
**Data source:** Appendix 4(60 countries, 2005)

```
Model 1: Pooled OLS, using 600 observations
Included 60 cross-sectional units        (Data for 60 countries)
Time-series length = 10                  (1965-2005, 5 years averages)
Dependent variable: log_YL_OutputPerCapita  (Y/L: Output per Capita)
-----------------------------------------------------------------------
| independent variable | coefficient | std. error | t-ratio | p-value |
-----------------------------------------------------------------------
const                   -0.0128322    0.752967     -0.01704   0.9864
ln_s_CapitalSavingsRate  0.467074     0.105969      4.408     1.24e-05 ***
ln_n+g+delta            -3.63336      0.266021    -13.66      3.81e-37 ***
ln_A_Technology          0.165705     0.033914      4.886     1.32e-06 ***
-----------------------------------------------------------------------

Mean dependent var   9.358455      S.D. dependent var   1.046538
Sum squared resid    417.2885      S.E. of regression   0.836749
R-squared            0.363938      Adjusted R-squared   0.360736
F(3, 596)            113.6718      P-value(F)           3.27e-58
Log-likelihood      -742.4176      Akaike criterion     1492.835
Schwarz criterion    1510.423      Hannan-Quinn         1499.682
rho                  0.900278      Durbin-Watson        0.130696
```

The regression has an *R²* of **0.36** which is far from perfect, but it's a good start. The p values for all of our regressors are small enough to confirm the support of our hypothesis. Our base Schwarz criterion is *1510* and Akaike criterion is *1492*. Log-likelihood has initial value is of *-742*, Sum of squared residuals is *417* and Standard error of regression is *0.83*. Those are our base starting values in this simplest model, which we will compare against the next, more detailed models.

Now, we enhance the model, by add human capital (*H*) to the regression:

```
Model 2: Pooled OLS, using 600 observations
Included 60 cross-sectional units        (Data for 60 countries)
Time-series length = 10                  (1965-2005, 5 years averages)
Dependent variable: log_YL_OutputPerCapita  (Y/L: Output per Capita)
--------------------------------------------------------------------------
|   independent variable   | coefficient | std. error | t-ratio |  p-value   |
--------------------------------------------------------------------------
const                           1.97318      0.640023     3.083    0.0021   ***
ln_s_CapitalSavingsRate         0.20869      0.089832     2.323    0.0205   **
ln_n+g+delta                   -2.18130      0.239397    -9.112    1.23e-18 ***
ln_HumanK_SecondaryEducation    0.97252      0.060145    16.170    5.13e-49 ***
ln_A_Technology                 0.08940      0.028682     3.117    0.0019   ***
--------------------------------------------------------------------------
Mean dependent var   9.358455      S.D. dependent var   1.046538
Sum squared resid    289.9012      S.E. of regression   0.698018
R-squared            0.558111      Adjusted R-squared   0.555140
F(4, 595)            187.8730      P-value(F)           5.1e-104
Log-likelihood      -633.1463      Akaike criterion     1276.293
Schwarz criterion    1298.277      Hannan-Quinn         1284.851
rho                  0.950464      Durbin-Watson        0.091270
```

Both Akaike and Schwarz criterions are with lower values, indicating that we have moved towards a better model. Log-likelihood has also improved. $R^2$ has improved to *0.55*. Sum of squared residuals and Standard error of regression has diminished as well.

As predicted, the residual of technology (*A*) has been split to human capital, acknowledging the link between technology (*A*) and human capital (*H*). In both models 1 and 2, technology function scales correctly with both classical and augmented growth functions.

*Experiment 3:* **Determining drivers of growth: Dynamic technology model with patents per capita**

Let's try to determine now which variables are exactly the drivers of growth. At this point, we have concluded technology (*A*) definitely helps improve output per capita(*Y/L*). However, technology (*A*) is calculated by taking a lot of factors. First, let's try to extend the previous model by including rate of patent discovery per capita (*Pa*) influencing output per capita (*Y/L*), both with and without human capital (*H*) considered:

Without human capital (*H*):

```
Model 3: Pooled OLS, using 600 observations
Included 60 cross-sectional units        (Data for 60 countries)
Time-series length = 10                  (1965-2005, 5 years averages)
Dependent variable: log_YL_OutputPerCapita  (Y/L: Output per Capita)
-----------------------------------------------------------------------------
| independent variable | coefficient | std. error | t-ratio |   p-value   |
-----------------------------------------------------------------------------
const                    10.969700     0.7400550    14.820    1.52e-42 ***
ln_s_CapitalSavingsRate   0.234235     0.0788010     2.972    0.0031   ***
ln_n+g+delta             -0.771498     0.2316930    -3.330    0.0009   ***
ln_Pc_IdeasPerCapita      0.322709     0.0138282    23.340    4.67e-86 ***
-----------------------------------------------------------------------------
Mean dependent var    9.358455     S.D. dependent var    1.046538
Sum squared resid     226.7770     S.E. of regression    0.616845
R-squared             0.654330     Adjusted R-squared    0.652590
F(3, 596)             376.0620     P-value(F)            5.2e-137
Log-likelihood       -559.4744     Akaike criterion      1126.949
Schwarz criterion     1144.536     Hannan-Quinn          1133.795
rho                   0.906242     Durbin-Watson         0.138761
```

With human capital (*H*) applied:

```
Model 4: Pooled OLS, using 600 observations
Included 60 cross-sectional units        (Data for 60 countries)
Time-series length = 10                  (1965-2005, 5 years averages)
Dependent variable: log_YL_OutputPerCapita  (Y/L: Output per Capita)
-----------------------------------------------------------------------------
| independent variable | coefficient | std. error | t-ratio |   p-value   |
-----------------------------------------------------------------------------
const                    9.84017      0.6940070    14.18     1.59e-39 ***
ln_s_CapitalSavingsRate  0.135369     0.0735828     1.84     0.0663   *
ln_n+g+delta            -0.527364     0.2157800    -2.44     0.0148   **
ln_HK_SecondaryEducation 0.551635     0.0548973    10.05     4.76e-22 ***
ln_Pc_IdeasPerCapita     0.254623     0.0144797    17.58     4.71e-56 ***
-----------------------------------------------------------------------------
Mean dependent var    9.358455     S.D. dependent var    1.046538
Sum squared resid     193.8761     S.E. of regression    0.570826
R-squared             0.704480     Adjusted R-squared    0.702493
F(4, 595)             354.5996     P-value(F)            6.7e-156
Log-likelihood       -512.4499     Akaike criterion      1034.900
Schwarz criterion     1056.885     Hannan-Quinn          1043.458
rho                   0.932169     Durbin-Watson         0.109587
```

Log-likelihood is improved. Sums of squared residuals and standard error of regression continue to diminish in both models. Akaike and Schwarz criterions indicate in both cases a better model, by their lower values. We notice a dramatic improvement with the new model in $R^2$ (*0.65* with and *0.70* without human capital (*H*)). All values indicate that we are moving towards a better model. However, the real interesting find here is with capital savings rate (*s*). After including human capital, it turns out savings rate start to be less statistically significant (*t* and *p* values), while the significance of it is split between human capital (*H*) and ideas per capita (*Pc*).

*Experiment 4:* **Determining growth drivers: Full model test**

This time I apply all variables (ideas, incentives, population growth rate, human capital level) in order to determine how the full model behaves against real data.

```
Model 5: Pooled OLS, using 117 observations
Included 60 cross-sectional units                (Data for 60 countries)
Time-series length: varying (minimum 1, maximum 3) (1996-2005)
Dependent variable: log_YL_OutputPerCapita        (Y/L: Output per Capita)
---------------------------------------------------------------------------
| independent variable | coefficient | std. error | t-ratio |  p-value   |
---------------------------------------------------------------------------
const                    7.6387500     1.606070      4.756    5.97e-06 ***
ln_s_CapitalSavingsRate  0.1952990     0.202492      0.965    0.3369
ln_n+g+delta            -0.6919760     0.485635     -1.425    0.1570
ln_HK_SecondaryEducation 1.4085900     0.268841      5.240    7.74e-07 ***
ln_IdeasPerCapita        0.0980516     0.050560      1.939    0.0550   *
ln_GDP_%_R&D             0.2785370     0.080770      3.449    0.0008  ***
---------------------------------------------------------------------------
Mean dependent var   9.774251     S.D. dependent var   1.026764
Sum squared resid    28.60124     S.E. of regression   0.507611
R-squared            0.766124     Adjusted R-squared   0.755589
F(5, 111)            72.72203     P-value(F)           2.07e-33
Log-likelihood      -83.60546     Akaike criterion     179.2109
Schwarz criterion    195.7840     Hannan-Quinn         185.9394
```

The assessment results are quite astonishing at this point with a $R^2$ of *0.75*. Sum of squared residuals is at all time low of *28* and Standard error of regression is also lowest so far: *0.50*. Schwarz and Akaike criterions are as low as *195.21* and *179.21*. This is indeed by far our best suggested model. It must be also noted that statistical significance (*p* value) of capital savings rate (*s*) has greatly diminished, confirming the notion of the original Solow model, that it cannot boosted endlessly to increase the output of the model, however unlike technology (*A*).

At this point both population growth rates and capital savings rates are becoming overshadowed by research incentives, by being at the edge of statistical insignificance. Another find is that the importance of ideas per capita is also slightly diminishing[26]. It is interesting to see that actually incentives for research

---

[26] I have incurred this in my other tests too. Ideas per capita start to diminish as a criteria of technological advancement as more detailed regressors are added. I wanted to test this also with percentage of labor employed in

play a significant role for growth (refer to **Appendix 10**). Human capital (*H*) accumulation also seems to contribute to growth quite well, despite the depreciation. This is not conflicting with my notion for depreciation since we are using real life data (with highest data of *H=13.10* (*ln(H)=2.5*) for year 2010 in USA) with depreciation applied to it it's *Hd=1.125*) which means that the data has experienced very little to no depreciation rate (since average 13 years of education means getting at most a 2-years college degree (or dropout), which is far from perfect, especially for the needs high-value added industries. Therefore, we still have far to go by improving our human capital (*H*) globally.

---

research, however the data available is only for 38 countries since 1996 on this indicator, with a lot of missing reports. My explanation for this insignificance is that not all ideas are actually put into production, and in some countries it is mostly big corporations filing patents. Thus, we need an even better measurement of ideas.

*Experiment 5:* **Determining critical growth values for human capital, R&D spending of GDP and ideas per capita by clusterization.**

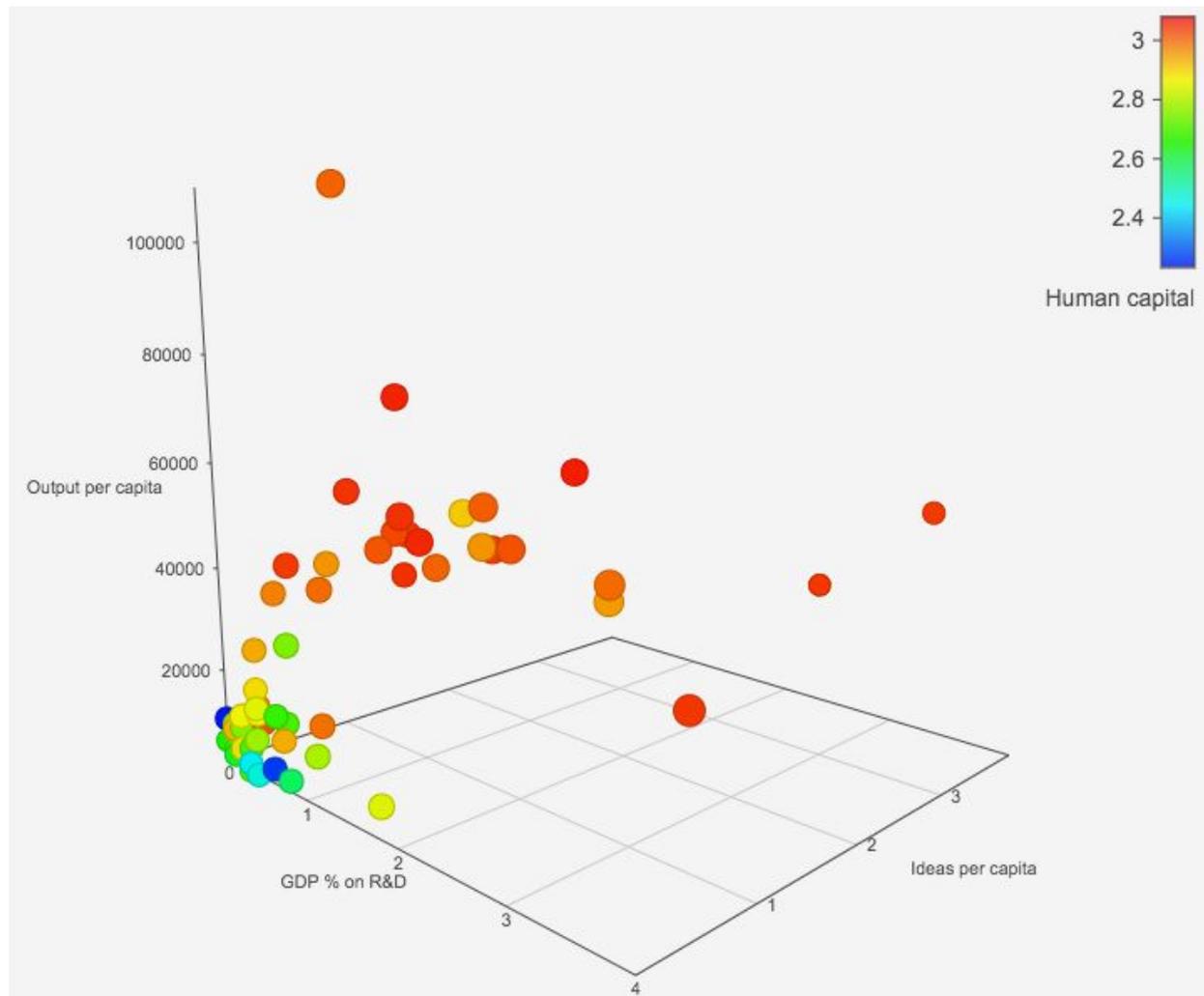

**Graph 7:** Output per capita vs combined technology indicators (Ideas per capita, GDP spending on research and human capital level). Each bubble is one country. Warmer colors indicate higher levels of human capital.
**Data source:** Appendix 3 (60 countries, 2005)

Here I use 4-dimensional plotting in order to better illustrate our the relationship between research incentives, ideas per capita, output per capita and human capital.

It turns out, poor states cannot benefit properly from technology (*A*), unless a critical mass of human capital (*H*) is reached, which would be able to grasp it. This is why, for a poor state it makes more sense first to advance in capital savings rate (*s*), regulate its population growth rates ***n*** (so no exponential growth or population extinction occurs) and increase human capital *H* (which has a diminishing returns cap), before focusing on research and innovation (*A*). We can notice the significant gap of human capital (*H*) between cutting-edge and catching-up growth. Romer (1990) claims that it is crucial to have a large human capital (*H*) for growth. While this is true, I would like to extend this notion by saying that for growth, human capital is needed, together with sufficient incentives for carrying out research and efficient institutions, with the latter being the true unexplained residual for me. The experiment reveals output per capita (*Y*) is boosted significantly (for future periods) when research is incentivized. This also explains in ancient history why development has started to occur only after some of the capital was saved from immediate consumption, and instead invested in education and technology.

*Experiment 6:* **Detecting data anomalies in the model via machine learning**
In this last experiment I test whether nations with similar technology levels (*A*) also have similar growth rates. Here I show how some machine learning algorithms such as K-means can be used to get meaningful subsets of data with similar characteristics and test them for anomalies. The data set is again with 60 countries and I clusterize the data by human capital (*H*), GDP spending on research and patent applications per capita:

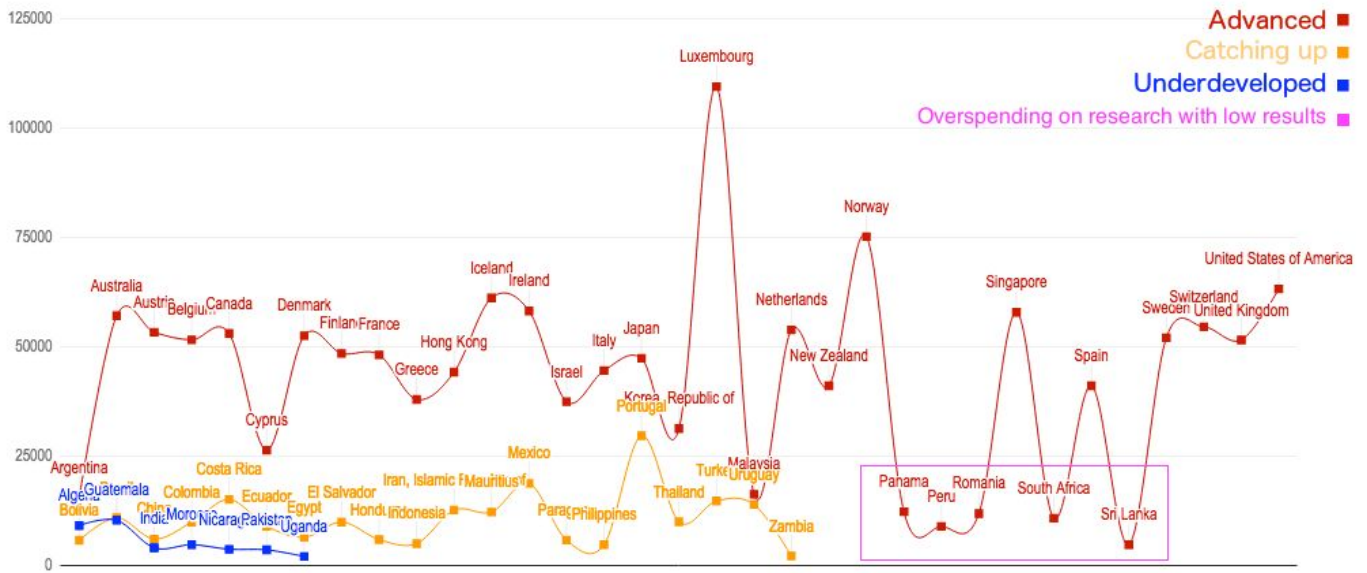

**Graph 8:** Clusterization of countries by technology (*A*), plotted by output per capita (*Y*).
**Data source:** Appendix 5

The cluster parameter is *k=3* (in order to split data to 3 clusters of similarity via technology). The results confirm the theory of divergence of cutting-edge and catching-up growth, together with the existence of divergent group. However it also helps us spot anomalies and "hidden groups of data". When K-means adds a member to a cluster, which is significantly away from the other cluster members, this means that the parameter must be increased, since more subclusters exists in the data set, and this member is an anomaly for the cluster.

We see that that the algorithm has placed in advanced technologies Panama, Peru, Romania, South Africa and Sri Lanka, which appear to have a decent investments in research (*A*), however, their output per capita is quite low. These can be countries which are either investing too much in research for their growth stage (catching-up), or countries which suffer from corruption or inefficient institutions (so their spending on A is ineffective to output per capita). For instance, out of 60 countries, Romania has the 14th most educated human capital in the list (***9.98 years***) at 26th place by patent discovery, but it's spending on research is at 42th (***0.4%***) with output per capita at 37th. Another anomaly is Luxembourg, which has a disproportionately high output per research. If we clusterize with *k=4*, the aforementioned countries are split to a separate cluster of data, by including other transitional countries with similar issues (see **Appendix 6**).

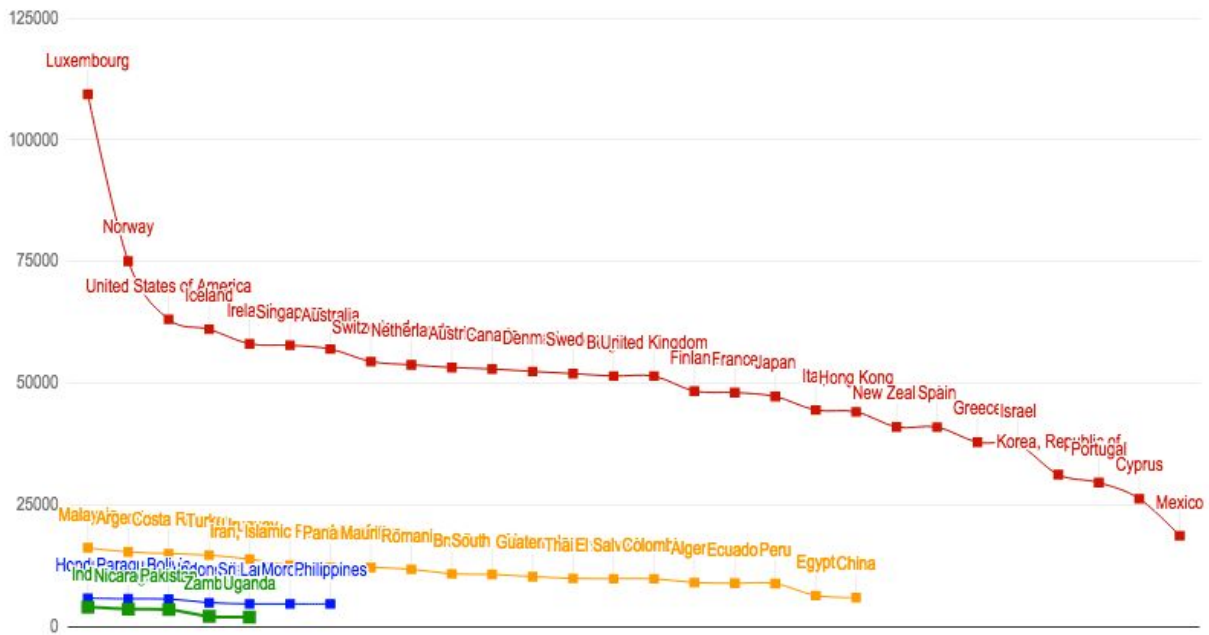

**Graph 9:** Clusterization of countries by technology (*A*), plotted by output per capita (*Y*). Parameter k=4
**Data source:** Appendix 6

## Conclusion

My stance is that technology is created by people, using time and resources, so using this links a model can be achieved. I wanted to prove that technology is affected endogenously, and it is too important and interconnected to be left out as exogenous constant.

**Closing words on the Solow model's issues**

The classic Solow model tells us that capital accumulation cannot explain long run economic growth. The human-capital augmented version attributes this to human capital accumulation. In the present paper, I have modeled the link between human capital and technology. However, still, this model does not answer the question why in some countries with the right technology, human capital and physical capital the output per capita can still stagnate (I have provided a detailed example about Japan in **Appendix 1**).

The reason for this is the presence of other variables, which are still missing in the production function model. And while Sala-i-Martin[27] is pointing out those which he found to be statistically significant, the majority of them are still not considered in growth functions. As noted by Paul Romer[28], currently there is a trend of retiring from generalized models (the larger the model, the more data is being lost, due to the use of models) in favor of smaller specific ones.

The endogenous growth model I propose still has a residual which needs to be further examined. I speak about the local political factors and regulations which Sala-i-Martin has found to be statistically significant[29]. During my research I have stumbled upon only one paper which tries to address institutions in the Solow model, however, without empirical evidence[30].

Mathematics must not be abused: it allows us to model processes, but with loss of details. We need to find reliable ways of abstraction in order to model the role of institutions for collaboration and government policies, which are multi-vector compound variables. My sincere belief is this can - and should be - done with absolute data, not with questionnaires and ratings ("spending a month in a country and rating its institutions").

I consider these to be the important missing link and the real "residual" in our detailed growth model.

---

# Appendixes

**1. Case study: The duality of productivity in Japan**

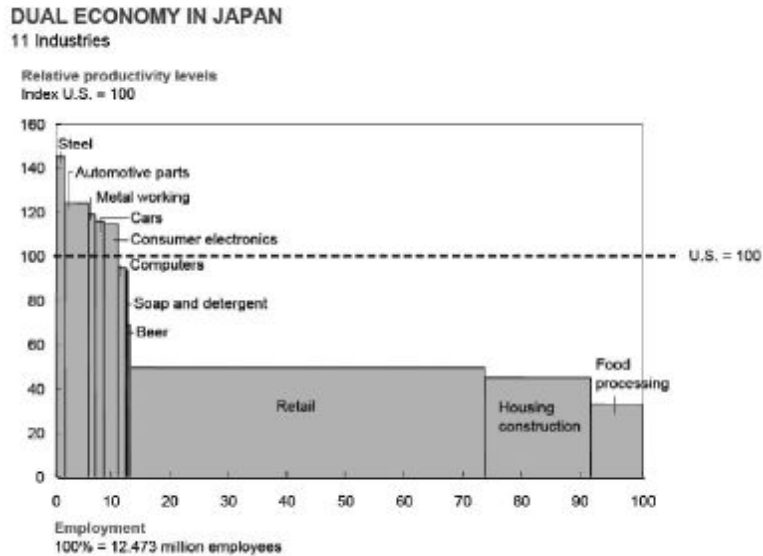

**Data source:** https://www.justice.gov/atr/speech/maximizing-welfare-through-technological-innovation

I would like also to point, that I do differentiate between Human Capital and Productivity since productivity is a function of first (or at worst, highly correlated to human capital). And while a country can have a very high level of technology, patents and research spending, they can still fail to benefit from it efficiently.

Let us take a look at Japan's productivity levels. Japan has very productive manufacturing sector, which supplies *~12%* of the employment, combined with a huge low productivity retail sector (*~55%* of employment). And this is a case, where even our endogenously augmented model would fail, as we are not measuring the microeconomics. The retail market in Japan is dominated by local "moms and pops shops", since it is restricted to local retailers, limiting competitiveness only to domestic level, while, for example, the US retailers are competing on the rough global markets[31].

While Japan has an exceptional human capital (*H*) and one of the highest rates of technology (*A*), they still combined it inefficiently, due to various regulations. This means that neither technology (*A*) or human capital (*H*) can be modeled to measure labor productivity. In order to measure productivity, or the lack of it, we need to consider also the local factors conditions.

---

[31] Barnett, T. (2007). *Maximizing welfare through technological innovation*. Presentation to the George Mason University Law Review. Washington, DC., [online] Available at: https://www.justice.gov/atr/speech/maximizing-welfare-through-technological-innovation [Accessed 28 Jul. 2017]

**2. Data sources**

1. **Patent applications by country residents:** Worldbank

2. **GDP percentage spent on research by country:** Worldbank

3. **Amount of researchers by country:** Worldbank

4. **Amount of technicians engaged in research by country:** Worldbank

5. **PENN:** Penn World Tables v7.1 (http://www.rug.nl/ggdc/productivity/pwt/pwt-releases/pwt-7.1)

6. **Development Indicators:** Worldbank

7. **Barro-Lee Data on Education** (http://www.barrolee.com/data/dataexp.htm)

**3. Built data set:** Sample data

**4. Built data set:** Real world data (merged and cleansed):

**5. Built data set:** Clusterization via K-means (filtered from Real world data) with parameter k=3

**6. Built data set:** Clusterization via K-means (filtered from Real world data) with parameter k=4

**7. Graphs:** Break-even point of growth function with technology augmentation (*A*):

| Comparison of results: Growth with and without research | | | Solow with human capital (H) | | | | Solow with human capital | | |
|---|---|---|---|---|---|---|---|---|---|
| Period | Y (No R&D) | Y (R&D) | Y (R&D) / Y: R&D effectiveness (scalar) | e | A | L | K | L (R&D) | K (R&D) | A R&D |
| 44 | 2102.5 | 2057.0 | 0.978 | 1.086 | 1 | 44 | 44 | 43.99 | 42.24 | 1.02 |
| 45 | 2203.2 | 2157.4 | 0.979 | 1.088 | 1 | 45 | 45 | 44.99 | 43.20 | 1.02 |
| 46 | 2306.4 | 2260.5 | 0.980 | 1.090 | 1 | 46 | 46 | 45.99 | 44.16 | 1.02 |
| 47 | 2412.2 | 2366.3 | 0.981 | 1.092 | 1 | 47 | 47 | 46.99 | 45.12 | 1.02 |
| 48 | 2520.6 | 2474.9 | 0.982 | 1.094 | 1 | 48 | 48 | 47.99 | 46.08 | 1.02 |
| 49 | 2631.5 | 2586.2 | 0.983 | 1.096 | 1 | 49 | 49 | 48.99 | 47.04 | 1.02 |
| 50 | 2745.0 | 2700.4 | 0.984 | 1.098 | 1 | 50 | 50 | 49.99 | 48.00 | 1.03 |
| 51 | 2861.1 | 2817.4 | 0.985 | 1.100 | 1 | 51 | 51 | 50.99 | 48.96 | 1.03 |
| 52 | 2979.8 | 2937.2 | 0.986 | 1.102 | 1 | 52 | 52 | 51.99 | 49.92 | 1.03 |
| 53 | 3101.1 | 3060.0 | 0.987 | 1.104 | 1 | 53 | 53 | 52.99 | 50.88 | 1.03 |
| 54 | 3225.1 | 3185.6 | 0.988 | 1.106 | 1 | 54 | 54 | 53.99 | 51.84 | 1.03 |
| 55 | 3351.7 | 3314.1 | 0.989 | 1.108 | 1 | 55 | 55 | 54.99 | 52.80 | 1.03 |
| 56 | 3481.0 | 3445.7 | 0.990 | 1.110 | 1 | 56 | 56 | 55.99 | 53.76 | 1.03 |
| 57 | 3612.9 | 3580.2 | 0.991 | 1.112 | 1 | 57 | 57 | 56.99 | 54.72 | 1.03 |
| 58 | 3747.5 | 3717.7 | 0.992 | 1.114 | 1 | 58 | 58 | 57.99 | 55.68 | 1.03 |
| 59 | 3884.8 | 3858.3 | 0.993 | 1.116 | 1 | 59 | 59 | 58.99 | 56.64 | 1.03 |
| 60 | 4024.8 | 4001.9 | 0.994 | 1.118 | 1 | 60 | 60 | 59.99 | 57.60 | 1.04 |
| 61 | 4167.5 | 4148.7 | 0.995 | 1.120 | 1 | 61 | 61 | 60.98 | 58.56 | 1.04 |
| 62 | 4313.0 | 4298.5 | 0.997 | 1.122 | 1 | 62 | 62 | 61.98 | 59.52 | 1.04 |
| 63 | 4461.2 | 4451.6 | 0.998 | 1.124 | 1 | 63 | 63 | 62.98 | 60.48 | 1.04 |
| 64 | 4612.1 | 4607.8 | 0.999 | 1.126 | 1 | 64 | 64 | 63.98 | 61.44 | 1.04 |
| 65 | 4765.8 | 4767.3 | 1.000 | 1.128 | 1 | 65 | 65 | 64.98 | 62.40 | 1.04 |
| 66 | 4922.3 | 4930.0 | 1.002 | 1.130 | 1 | 66 | 66 | 65.98 | 63.36 | 1.04 |
| 67 | 5081.5 | 5096.0 | 1.003 | 1.132 | 1 | 67 | 67 | 66.98 | 64.32 | 1.04 |
| 68 | 5243.6 | 5265.3 | 1.004 | 1.134 | 1 | 68 | 68 | 67.98 | 65.28 | 1.05 |
| 69 | 5408.5 | 5438.0 | 1.005 | 1.136 | 1 | 69 | 69 | 68.98 | 66.24 | 1.05 |

**8. Graph:** Output per capita (Y/L) vs Ideas per capita (Pc) comparison

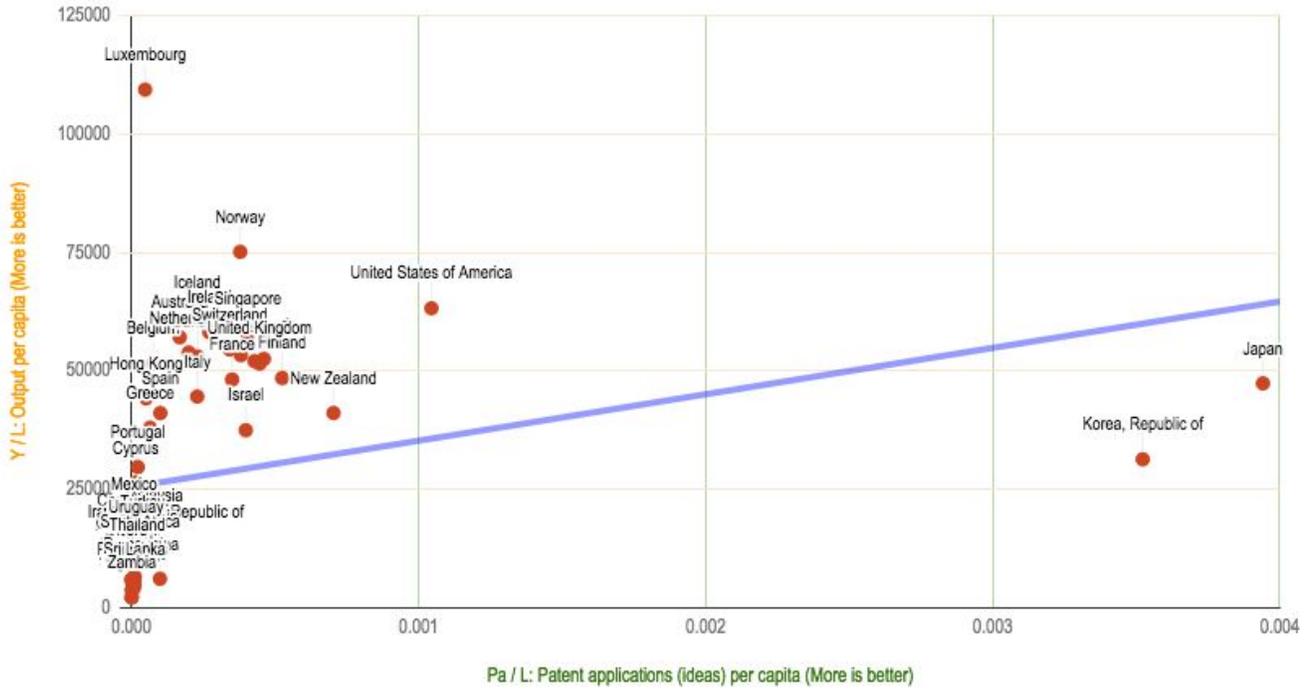

**9. Graph:** Output per capita (Y/L) vs Human capital (with depreciation rate applied):

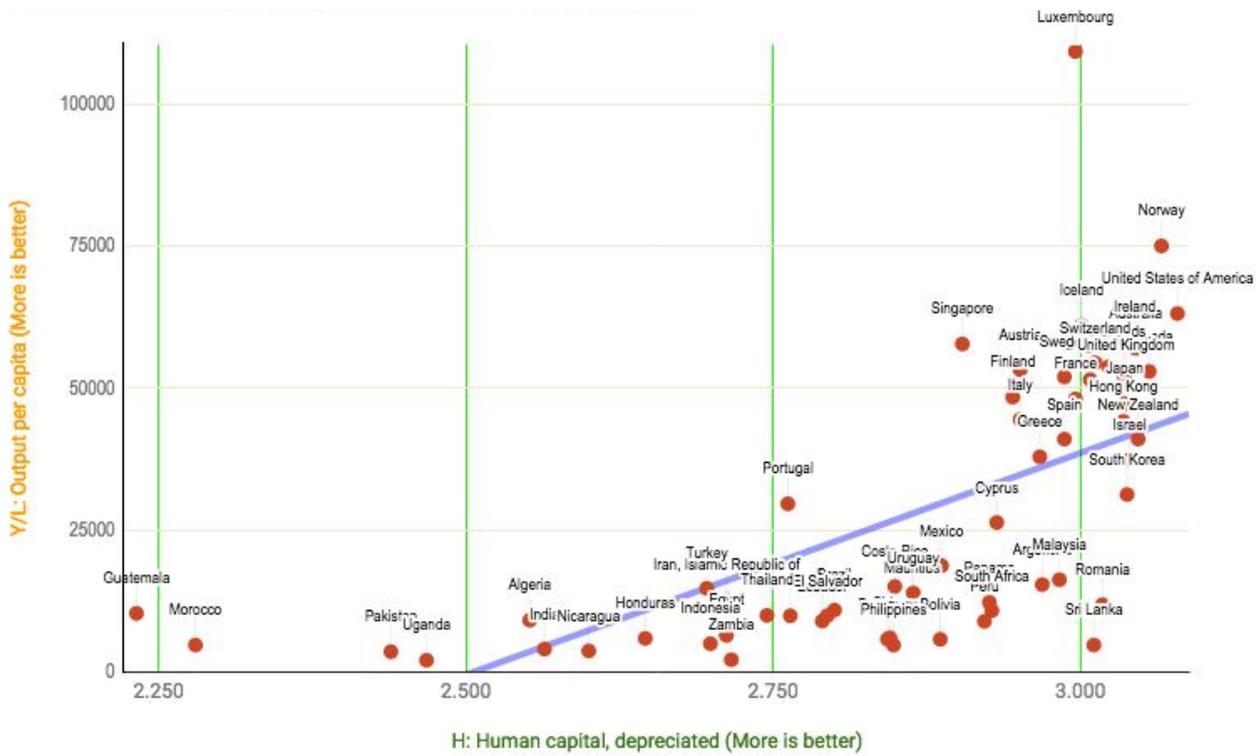

**10. Graph:** Output per capita (Y/L) vs Incentives (Percent of GDP spent on research):

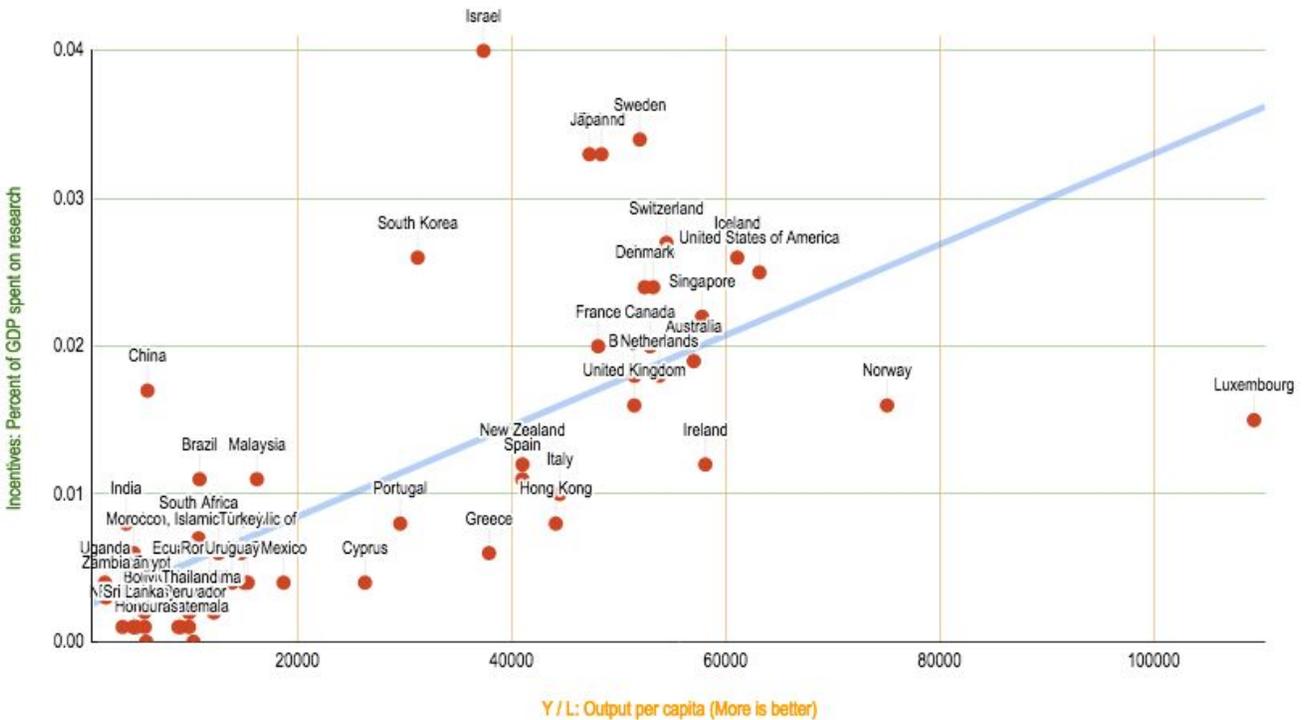